\documentclass[11pt,dvips]{article}
\usepackage{graphicx}
\usepackage{epsfig,times} 
\usepackage{picinpar}
\usepackage{wrapfig}
\usepackage{floatflt}
\makeatletter
\renewcommand{\section}{\@startsection{section}{1}{0in}
	{0.4\baselineskip}{0.1\baselineskip}{\Large\bf}}
\renewcommand{\subsection}{\@startsection{subsection}{2}{0in}
	{0.25\baselineskip}{-\baselineskip}{\large\bf}}
\renewcommand{\subsubsection}{\@startsection{subsubsection}{3}{0in}
	{0.1\baselineskip}{-\baselineskip}{\normalsize\bf}}
\makeatother
\begin{document}

%
\thispagestyle{myheadings}
%
\markright{OG 2.2.20}
\begin{center}
%
{\LARGE \bf Search for TeV Gamma Rays from the SNR RXJ1713.7$-$3946}
\end{center}

\begin{center}
%
%

{\bf
H.~Muraishi\footnote[1]{
Faculty of Science, Ibaraki University, 
   Mito, Ibaraki 310-8521, Japan, 
$^{2}$Department of Physics, Tokyo Institute of Technology, 
        Meguro, Tokyo 152-8551, Japan, 
$^{3}$Institute for Cosmic Ray Research, University of Tokyo,
     Tanashi, Tokyo 188-8502, Japan, 
$^{4}$Department of Physics and Mathematical Physics, University of 
   Adelaide, South Australia 5005, Australia, 
$^{5}$Institute of Space and Astronautical Science,
   Sagamihara, Kanagawa 229-8510, Japan, 
$^{6}$Department of Physics, Yamagata University, 
Yamagata 990-8560, Japan, 
$^{7}$Faculty of Management Information, Yamanashi Gakuin University,  Kofu, 
Yamanashi 400-8575, Japan, 
$^{8}$Department of Physics, Tokai University, 
 Hiratsuka, Kanagawa 259-1292, Japan, 
$^{9}$STE Laboratory, Nagoya University,
   Nagoya, Aichi 464-860, Japan, 
$^{10}$National Astronomical Observatory, Tokyo 181-8588, Japan, 
$^{11}$LPNHE, Ecole Polytechnique. Palaiseau CEDEX 91128, France,
$^{12}$Computational Science Laboratory, Institute of Physical and Chemical
   Research, Wako, Saitama 351-0198, Japan,
$^{13}$Faculty of Engineering, Kanagawa University,
 Yokohama, Kanagawa 221-8686, Japan
}, 
T.~Tanimori$^2$, S.~Yanagita$^{1}$, T.~Yoshida$^{1}$, T.~Kifune$^3$,
S.A.~Dazeley$^4$,
P.G.~Edwards$^5$,
S.~Gunji$^6$, S.~Hara$^2$,  
T.~Hara$^7$, J.~Jinbo$^8$, 
A.~Kawachi$^3$,
H.~Kubo$^2$, 
J.~Kushida$^2$, Y.~Matsubara$^9$, 
Y.~Mizumoto$^{10}$, 
M.~Mori$^3$,
M.~Moriya$^2$, 
Y.~Muraki$^9$, 
T.~Naito$^7$, K.~Nishijima$^8$, 
J.R.~Patterson$^4$, M.D.~Roberts$^3$, 
G.P.~Rowell$^3$, T.~Sako$^{9,11}$, 
K.~Sakurazawa$^2$, Y.~Sato$^3$, R.~Susukita$^{12}$, 
T.~Tamura$^{13}$,  
T.~Yoshikoshi$^3$, 
A.~Yuki$^9$  
}
\end{center}

\begin{center}
{\large \bf Abstract\\}
\end{center}
\vspace{-0.5ex}
%
%
The shell type SNR RXJ1713.7$-$3946 is a new SNR discovered by
the ROSAT all sky survey. Recently, strong non-thermal X-ray
emission from the northwest part of the remnant was detected by
the ASCA satellite. This synchrotron X-ray emission
strongly suggests the existence of electrons with energies
up to hundreds of TeV in the remnant. This SNR is, therefore,
a good candidate TeV gamma ray source,
due to the Inverse Compton scattering of the Cosmic Microwave
Background Radiation by the shock accelerated
ultra-relativistic electrons, as seen in SN1006. In this paper,
we report a preliminary result of TeV gamma-ray observations of
the SNR RXJ1713.7$-$3946 by the CANGAROO 3.8m telescope 
at Woomera, South Australia.
%

\vspace{1ex}

%
%
\section{Introduction:}
\label{intro.sec}

The recent result of the observation of type Ia SNR SN1006 by ASCA
demonstrates that there exist electrons with energies up to $\sim$ 100
TeV which are accelerated within the shock front of the remnant
(Koyama et al. 1995). This finding is strong evidence for the SNR origin
of cosmic rays. The existence of electrons with energies around 100
TeV is demonstrated more directly by the CANGAROO observation of
TeV gamma-rays from the northeast rim of SN1006, which coincides with
the region of maximum flux in the 2--10 keV band of the ASCA data
(Tanimori et al. 1998). This TeV gamma-ray emission was explained
as the 2.7 K cosmic background photons up-scattered by electrons with
energies up to $\sim$ 100 TeV by the IC process and allowed, 
together with the observation of non-thermal
radio and X-ray emission,
the physical parameters of the remnant, such as the
magnetic field strength,  to be estimated
(Mastichiadis 1996 ; Mastichiadis \& De Jager 1996 ;
Yoshida \& Yanagita 1997 ; Pohl 1996).

Recently, a new shell type SNR, RX\,J1713.7$-$3946, was discovered in
the ROSAT all-sky survey (Pfeffermann et al. 1996). The remnant has a
slightly elliptical shape with a maximum extent of $\sim$ 70$\prime$. The
total thermal X-ray flux from the whole remnant was estimated as
$\sim$ 4.4 $\times$ 10$^{-10}$ erg cm$^{-2}$ s$^{-1}$ in the 0.1 -- 2.4
keV energy band, ranking the remnant among the brightest galactic
supernova remnants. The remnant was put at a distance of 1.1kpc, with
an age of only $\sim$ 2100 years from the Sedov solution. Subsequent
observations of this new remnant by the ASCA Galactic Plane Survey
revealed strong non-thermal hard X-ray emission from the northwest
(NW) rim of the remnant that is three times higher than that from
SN1006 (Koyama et al. 1997). It is suspected that the X-rays from
RX\,J1713.7$-$3946 are dominated by this non-thermal emission from the
NW rim. The dominance of non-thermal emission from the shell is
reminiscent of SN1006. Koyama et al. proposed from the global
similarity of the new remnant to SN1006 -- in its shell type
morphology, the non-thermal nature of the X-ray emission, and apparent
lack of central engine like a pulsar -- that RX\,J1713.7$-$3946 is the
second example, after SN1006, of synchrotron X-ray radiation from a
shell SNR.

If this scenario is correct, there is the possibility of
observing TeV gamma-ray emission from the new remnant as seen in
SN1006. With this motivation, we have observed RX\,J1713.7$-$3946 with the
CANGAROO imaging TeV gamma-ray telescope in 1998. Here we report a
preliminary result of these observations.

\section{Instrument and Observation:}
\label{format.sec}
The CANGAROO 3.8m imaging TeV gamma-ray telescope is located near
Woomera, South Australia (136$^{\circ}$47'E, 31$^{\circ}$06'S)
(Patterson \& Kifune 1992 ; Hara et al. 1993). A high resolution
camera of 256 photomultiplier tubes (Hamamatsu R2248) is installed in
the focal plane. The field of view of each tube is about
0$^{\circ}$.12 $\times$ 0$^{\circ}$.12, and the total field view of
the camera is about 3$^{\circ}$. The pointing accuracy of the
telescope is less than 0$^{\circ}$.02 from a study of the trajectory
of bright stars in the field of view (Yoshikoshi 1996). The
point-spread function (PSF) of the telescope is estimated to have a
standard deviation of 0$^{\circ}$.18 when fitted with a Gaussian
function assuming a point source. The sensitivity of our telescope
for an integral flux above 2 TeV is estimated as $\sim$ 2 $\times$ 10
$^{-12}$ photons cm$^{-2}$ s$^{-1}$ at zenith angle of 20$^{\circ}$
which is the average value for the observations of RX\,J1713.7$-$3946
from the CANGAROO telescope site.

RX\,J1713.7$-$3946 was observed from May to August
in 1998. During on-source observations, the center of the field of view
was pointed to the NW rim, which is the brightest region in the hard X-ray
band. The total observation time was 66 hours for on-source data and 64
hours for off-source data. After rejecting data affected by
cloud, a total of 42 hours of on- and off-source data remained for
this analysis.

\section{Analyses and Preliminary result:}
\label{additional.sec}

The imaging analysis of the data is based on the usual
parameterization of the elongated shape of the Cerenkov light image using
``width,''``length,''``concentration'' (shape), ``distance'' (location),
and the image orientation angle ``alpha'' (Hillas 1985 ; Weekes et al. 
1989 ; Reynolds et al. 1993). 
The values of these parameters differ from an image by gamma-rays to
that by cosmic rays and are utilized to discriminate gamma-ray events
from background cosmic ray events. The parameter ``alpha'' is most
efficient in this discrimination.
The application of this technique to data recorded with the CANGAROO
telescope has, to date, resulted in the detection of TeV gamma-rays
from PSR 1706$-$44 (Kifune et al. 1995), Crab pulsar/nebula (Tanimori
et al. 1994, 1998), Vela pulsar (Yoshikoshi et al. 1997) and SN1006
(Tanimori et al. 1998).

Figure 1 shows the resultant alpha distribution when we selected
gamma-ray like events with the criteria of
0.$^{\circ}$01$\le$width$\le$0.$^{\circ}$1,
0.$^{\circ}$1$\le$length$\le$0.$^{\circ}$4,
0.4$\le$concentration$\le$0.9 and
0.$^{\circ}$5$\le$distance$\le$1.$^{\circ}$2 with the NW rim of
RX\,J1713.7$-$3946 as the center of the field of view. The solid line
and the dashed line indicate the on-source and
off-source data respectively. Some excess can be seen at alpha $\sim$
0$^{\circ}$ for the on-source data. 
The fact that significance does not fall significantly with the wider
alpha acceptance, suggests the emitting region of TeV gamma-rays is
extended as in the case of SN1006. Although this preliminary analysis
yields a 5.0 $\sigma$ excess we require a more complete analysis and
confirming observations before RX\,J1713.7$-$3946 can be declared
an established TeV gamma-ray source.

In order to examine if the TeV gamma-rays are emitted from extended
regions, the significances of events with alpha $\le$ 20$^{\circ}$ were
calculated at all grid points in 0.05$^{\circ}$ steps in the field of view. 
The resulting contour map of significances is shown in Figure 2, in
which the contours of the hard X-ray flux also are overlaid as solid
lines (Tomida 1999). The region which shows the emission of TeV
gamma-rays with relatively high significance seems to be extended and
to coincide with the ridge of the NW rim, which is the bright region in
hard X-rays as seen by ASCA.

ASCA reported that the non-thermal X-ray flux from RX\,J1713.7$-$3946 is
three times higher than that from SN1006 (Koyama et al. 1997). Accordingly it
may be interesting to compare the flux of TeV gamma-rays from
RX\,J1713.7$-$3946 with that from SN1006. If we estimate tentatively the
integral flux of TeV gamma-rays from the remnant of RX\,J1713.7$-$3946
based on the Monte Carlo simulation method by assuming the emission is
from a point source, the value is roughly 3 $\times$
10$^{-12}$ cm$^{-2}$ s$^{-1}$ ( $\ge$ 2 TeV ). This value is
comparable to that from SN1006. However it is premature to discuss
quantitatively the difference or the similarity between the two SNRs
due to the preliminary nature of our TeV gamma-ray result and also due
to an absence of detailed radio data for RX\,J1713.7$-$3946.

\section{Summary:}
We have observed a marginal excess emission of TeV gamma-rays from the
shell type SNR RX\,J1713.7$-$3946. The emitting region seems to be
extended and to coincide with the NW rim of the remnant bright in hard
X-rays as seen by ASCA. The SNR RX\,J1713.7$-$3946 is reminiscent of
SN1006 both in the synchrotron X-ray emissions from the shell of the
remnant and also in the TeV gamma-ray emissions from an extended region
coinciding with the region emitting the non-thermal X-rays. 
Further investigation of the TeV gamma-ray emission from RXJ1713.7$-$3946
will give us valuable information for a better understanding of
particle acceleration processes in SNRs.

%
%
%
%
%
%
\vspace{1ex}
\begin{center}
{\Large\bf References}
\end{center}
%
Hara, T., et al. 1993, Nucl. Inst. Meth. Phys. Res. A 332, 300\\
Hillas, A. M. 1985, 19th Int. Cosmic Ray Conf. (La Jolla) 3, 445\\
Kifune, T, et al. 1995, ApJL 438, L91-L94\\
Koyama, K., et al. 1995, Nature 378, 255\\
Koyama, K., et al. 1997, PASJ 49, L7\\
Mastichiadis, A. 1996, A\&A 305, L53\\
Mastichiadis, A. and de Jager.O.C. 1996, A\&A 311, L5\\
Patterson, J. R., and Kifune, T. 1992, Australian New Zealand Phys. 29, 58\\
Pfeffermann, E., Aschenbach B. 1996, in Roentgenstrahlung from the Universe, International Conference on X-ray Astronomy and Astrophysics, ed H.U. Zimmermann, J.E. Truemper, H. Yorke, MPE Report 263, P267\\
Pohl, M. 1996, A\&A 307, 57\\
Reynolds, P. T., et al. 1993, ApJ 404, 206\\
Tanimori, T, et al. 1994, ApJL 429, L61-L64\\
Tanimori, T, et al. 1998, ApJL 492, L33-L36\\
Tanimori, T, et al. 1998, ApJL 497, L25-L28 and Plate L2\\
Tomida, H. 1999, Ph. D. thesis, Kyoto University\\
Weekes, T.C., et al. 1989, ApJ 342, 379\\
Yoshida, T. and Yanagita, S. 1997, Proc. 2nd INTEGRAL Workshop `Transparent Universe' ESASP-382, 85\\
Yoshikoshi, T. 1996, Ph. D. thesis, Tokyo Institute of Technology\\
Yoshikoshi, T, et al. 1997, ApJL 487, L65-L68\\

\newpage
\begin{figure}[pt]
  \begin{center}
  \psfig{file=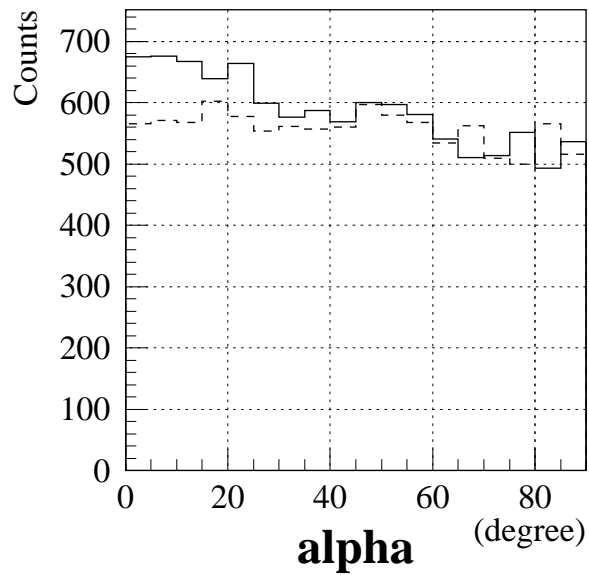,scale=1.0}
  \end{center}
\caption{Distributions of the image orientation angle ``alpha'' with respect to the direction of the NW rim, where the solid line is for on-source data and the dashed line is for off-source data.}
\end{figure}

\begin{figure}[b]
  \begin{center}
  \psfig{file=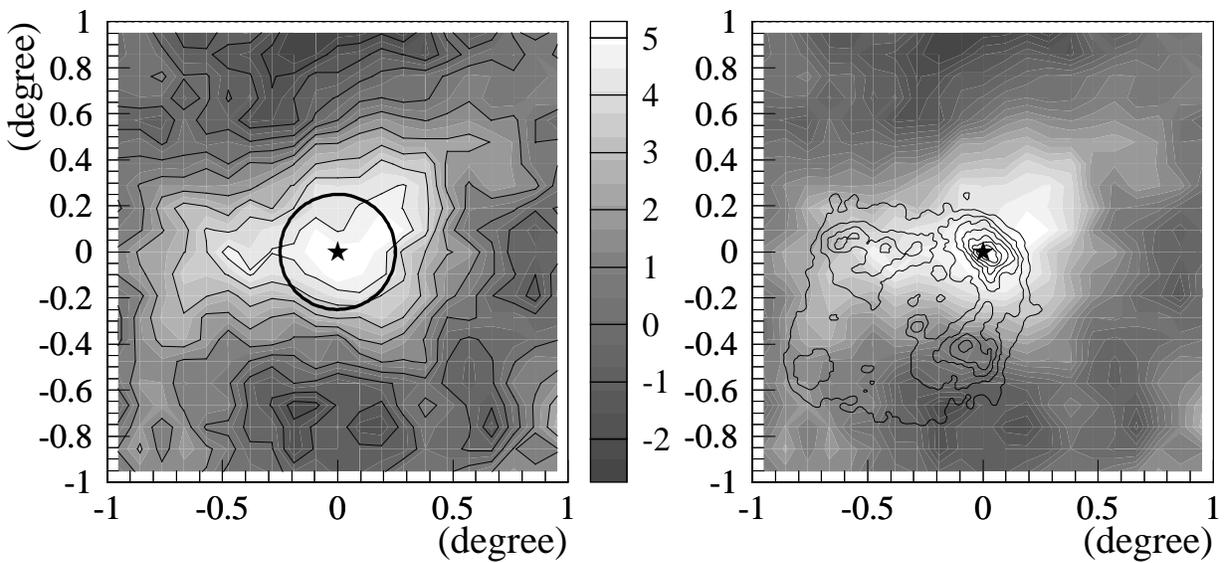,scale=1.0}
  \end{center}
\caption{Contour map of the statistical significance around the NW rim
of RX\,J1713.7$-$3946
plotted as a function of right ascension and declination; north
is up, and east is to the left. The contours of the hard X-ray flux
also are overlaid as solid lines (Tomida 1999) in the right-hand figure. 
The solid circle in the left figure is the area of the PSF of the
CANGAROO telescope assuming alpha $\le$ 20$^\circ$.}
\end{figure}

\end{document}